\documentstyle[12pt]{article}
\begin{document}
\baselineskip=6mm
\rightline{KEK-TH-416}
\rightline{KEK Preprint 94-127}
\rightline{KOBE-TH-94-02}
\vspace{2.0cm}
\centerline{{\large{\bf A Natural Explanation of All Solar Neutrino Data}}}
\par
\centerline{{\large{\bf by Resonant Spin-Flavor Precession Scenario}}}
\par
\par\bigskip
\par\bigskip
\par\bigskip
\par\bigskip
\par\bigskip
\centerline{{\bf C. S.  Lim} and {\bf H. Nunokawa}$^{\dagger}$} \par
\par\bigskip
\par\bigskip
\centerline{\sl Department of Physics,  Kobe University,  Nada,  Kobe 657,
Japan}
\par
\centerline{\sl Theory Group, KEK, Oho,  Tsukuba 305,
            Japan $^{\dagger}$} \par
\par\bigskip
\par\bigskip
\par\bigskip
\par\bigskip
\par\bigskip
\par\bigskip
\par\bigskip
\par\bigskip
\par\bigskip
\centerline{{\bf Abstract}}\par

It is emphasized that the $E_{\nu}$ (neutrino energy) spectrum of
the $\nu_e$ survival
probability in the Resonant Spin-Flavor Precession (RSFP) scenario
has very desirable shape to reconcile all existing
solar neutrino data.
As the result,  the RSFP scenario is shown to have rather broad allowed
region to reconcile the data in the ($B,\   \Delta m^2$) plane,  with
$B$ being the magnetic field strength inside the Sun.
The sensitivity of the allowed
region on the different choices of the $B$ profile,  and the time
variations of the solar neutrino event rates in the RSFP scenario are
also discussed in some detail.

\par\bigskip
\par\bigskip
\par\bigskip
\par\bigskip
\par\bigskip
\par\bigskip
\par\bigskip

\noindent October 1994
\newpage

\noindent
The solar neutrino problem seems to be almost unique remaining puzzle,
which cannot be resolved by the standard model of particle interactions,
and therefore requires some ``new physics". The most simple and elegant
particle physics solution to this problem by Mikheyev,  Smirnov,
and Wolfenstein (MSW) \cite{Mikheyev} relies on the resonant enhancement
 of neutrino oscillations inside the medium of the Sun. Another attractive
 solution was proposed by Okun,   Voloshin,  and Vysotsky (OVV) \cite{LOkun},
 which originally aimed to explain not only the deficit of the solar
neutrino event rates,  but also the possible anticorrelation of the solar
neutrino flux with the sunspot number,  claimed to exist in the Cl (Homestake)
experiment \cite{RDavis}. The scenario of OVV assumes that $\nu_e$
possesses a ``large" magnetic moment. The ``large" magnetic
moment,  when coupled to the strong transverse magnetic field inside
the Sun,  leads to a spin precession $\nu_{e_L} \rightarrow
\nu_{e_R}$,  yielding a sterile state,  $\nu_{e_R}$,  which escapes the
detection. Once the interaction of $\nu_{e_L}$ with
the solar matter is taken into account,  however,  the matter effect
causes an energy gap between $\nu_{e_L}$ and $\nu_{e_R}$ and tends
to prevent the spin precession. We should also notice that the extent
of spin precession has nothing to do with the neutrino energy $E_{\nu}$.
To account for all of the now available data on the three types of
solar neutrino experiments,  i.e. Cl (Homestake)\cite{RDavis,KLande},
$\nu e$ (Kamiokande)
\cite{KSHirata,YSuzuki},  and
Ga (SAGE \cite{SAGE,Nico} and GALLEX \cite{GALLEX})
experiments,  therefore seems to be a non-trivial task,
since these experiments have
sensitivities on different ranges of $E_{\nu}$ and are reporting different
$\nu_e$ deficit rates.

These problems can be evaded in a natural manner in the scenario of
Resonant Spin-Flavor Precession (RSFP for short)
\cite{WMarciano,Akhmedov}.
 The basic observation of the scenario is that the matter
effect,  which prevents the spin precession in the OVV scenario,  now
works to realize resonant enhancement of the spin-flavor precessions,
e.g. $\nu_{e_L} \rightarrow \nu_{\mu_R}$  for Dirac type precession,
$\nu_{e_L} \rightarrow \overline{\nu}_{\mu_R}$$ (= (\nu_{\mu_L})^C)$
for Majorana type precession. The deficit rate of $\nu_e$ now has
a $E_{\nu}$ dependence. (In the Majorana case the spin precession
$\nu_{e_L} \rightarrow \overline{\nu}_{e_R}$ is forbidden by CPT and
the spin-flavor precession via transition magnetic moment is an
inevitable choice.) The Majorana type RSFP,  which we will
consider in this paper,  has additional bonuses:
i) In the Majorana type transition the final state neutrino,
e.g. $\overline{\nu}_{\mu_R}$,  is no longer sterile and does
have a contribution to the event rate at Kamiokande.
This feature is desirable in trying to explain the higher event rate and
the milder time-dependence of the rate at Kamiokande compared with
those in Cl data \cite{MMori}. ii) Since the antineutrinos are trapped in
dense stars and since sterile neutrinos do not appear in the interaction
due to the transition moment,  the severe astrophysical constraint
coming from SN1987A
neutrino data \cite{Lattimer} and the cosmological constraint
on the number of neutrino species $N_{\nu} < 3.4$ \cite{MFukugita}
can be easily lifted.

Now that the data from all of the three types of solar neutrino experiments
are available,  the RSFP scenario is ready to be confronted with these data.
In particular,  the most recent data from the Ga experiments
\cite{SAGE,Nico,GALLEX}
show relatively high event rates,  and the result has been
argued to disfavor the RSFP,  since it contradicts with the large depletion
of the $pp$ neutrinos predicted in the interesting works in Ref.
\cite{RMohapatra},  which tried to reconcile the Cl
and $\nu e$ results (before the appearance of the Ga data),
paying attention to their different energy thresholds.

In this paper,  however,  the all different deficit rates of the three
types of experiments are shown to be simultaneously explained in a natural
manner in the framework of RSFP scenario. This controversy suggests
that the large
depletion of the $pp$ neutrino is not a genuine feature of RSFP,  but
rather may be due to the choice of parameters in the theory. We thus feel
it very important to make clear what is the genuine predictions of RFSP,
and which kinds of properties depend upon the specific choices of the
parameters,  before we derive some definite conclusion on the validity
of the scenario. In the present paper we therefore try to extract some
generic prediction of RSFP and to confront it with the observations.
As such a generic feature we focus on the specific $E_{\nu}$
dependence of the survival probability of $\nu_e$,
$P_{\nu} \equiv P(\nu_e \rightarrow \nu_e)$,  in RSFP.

We would like to emphasize that the $E_{\nu}$ dependence of $P_{\nu}$
in RSFP is exactly what we want in order to explain the different
event rates in three types of experiments,  and is quite different from the
$E_{\nu}$ dependence in MSW. Namely,  as is seen in Fig. 1,  which shows a
generic $E_{\nu}$ dependence in RSFP for suitable range of the
transverse magnetic field $B$,  the survival probability behaves as,
$P_{\nu} \sim 1$ for smaller $E_{\nu}$,  $P_{\nu} \sim 0$ for intermediate
$E_{\nu}$,  and $P_{\nu} \sim 1/2 $ for higher $E_{\nu}$. This behavior
is immediately realized to be very desirable to reconcile the reported
averaged deficit rates (observed event rates $/$ Standard Solar
Model predictions)
in Ga,  Cl,  and $\nu e$ experiments,  which are sensitive to
lower,  intermediate,  and higher energy solar neutrinos,  roughly
speaking:

\noindent Ga: $0.52 \pm 0.09$ (SAGE \cite{Nico}),
$0.60 \pm 0.09$ (GALLEX \cite{GALLEX})

\noindent Cl: $0.32 \pm 0.03 \  \cite{KLande}$

\noindent $\nu e$: $0.51 \pm 0.07 \ \cite{YSuzuki} $

\noindent  In fact,  in the present paper,  we will explicitly demonstrate
that,  owing to the specific $E_{\nu}$ dependence,  there exists a rather
broad range in $(B,\   \Delta m^2)$ plane of band shape,  which is an
``allowed region" in order to reconcile all of the four experimental
data. (This possibility was explicitly pointed out in
Ref. \cite{CLim}.) On the other hand,  in
MSW solution $P_{\nu}$ approaches to 1,  instead of 1/2,  as $E_{\nu}$
becomes very large.
Thus to get a reasonably good fit to the
observations in MSW,  $\theta$ (generation mixing angle) and $\Delta m^2$
have to be carefully chosen \cite{Hata}.
This is the reason why in MSW there remain
only two isolated rather restricted allowed regions in the
$(\theta,\  \Delta m^2)$ plane,  though this restriction at the same
time may enable us to pin down the correct set of parameters once the data
get further improved.

       After the results of Ga experiments were put forward,  there have
appeared some papers \cite{Akhmedov2,Plido},  which also claim
that RSFP scenario can reconcile the three kinds of experimental data.
(As for the argument in the ``hybrid",  MSW plus RSFP,  scenario
\cite{HMinakata},  see Ref \cite{HNunokawa}). In these papers the
importance of the careful choice of the profile (position dependence)
of the magnetic field $B$ has been emphasized. Since our main point is to
show that the all of the three deficit rates can be accounted for as a
consequence of the desirable $E_{\nu}$ dependence of $P_{\nu}$ in
RSFP, rather than the consequence of some special choices of $B$ profile, we
will be mainly concerned with the case with a constant $B$ throughout
the path of the solar neutrino propagation. Though the average deficit
rates of three types of experiments are naturally reconciled
for the constant $B$ for rather broad range of the parameter space,
we will further discuss how the allowed region is modified if $B$ has
different strengths in each of the convective
($ 0.7 \leq r/r_\odot \leq 1.0$ )
and the inner zones ($ r/r_\odot < 0.7$ )
of the Sun. While our main focus is on the average deficit rates,
we will also make some brief comments on the possible
time variations of the event rates,  since it was a original motivation
of the OVV and RSFP scenarios.

Since our main purpose in the present paper is to clarify to what
extent the specific features of RSFP,  especially its desirable $E_{\nu}$
dependence of the survival probability $P_{\nu}$,  are useful in order to
reconcile the three kinds of data,  we consider pure RSFP of Majorana type in
the 2 generation case,  $\nu_{e_L} \rightarrow \overline{\nu}_{\mu_R}$,
and neglect the generation mixing $\theta$. (We will not include the
interesting effect \cite{JVidal,Krastev} due to the twisting
of the magnetic field either.)

The time evolution of the system is governed by a Schr\"odinger-like
equation \cite{WMarciano};

\begin{equation}
i \frac{d}{dt}
\left(
\begin{array}{l}
\nu_{e_L} \\
\overline{\nu}_{\mu_R}
\end{array}
\right)
=
\left(
\begin{array}{cl}
a_{\nu_e} \,  \,   &   \mu B    \\
\mu B     \,  \,   &   \frac{\Delta m^2}{2 E_{\nu}} - a_{\nu_{\mu}}
\end{array}
\right)
\left(
\begin{array}{l}
\nu_{e_L}   \\
\overline{\nu}_{\mu_R}
\end{array}
\right)  ,
\end{equation}

\noindent where $\mu$ is the transition magnetic moment and the
matter effects of neutrinos are given as $a_{\nu_e} = \frac{G_F}
{\sqrt{2}}(2N_e - N_n)$ and
$a_{\nu_{\mu}} = \frac{G_F}{\sqrt{2}}(-N_n)$,  with $N_e$ and $N_n$
being electron and neutron number densities. In getting predictions on
the event rates in the three experiments we will numerically integrate
the above equation,  with the energy spectrum of each solar neutrino
source and the absorption cross sections for Cl and Ga experiments
being taken from Ref. \cite{JBahcall}. As for the $\nu e$ experiment,
the contribution from the $\overline{\nu}_{\mu_R} e$ scattering is
included
together with the suitable trigger efficiency for
the Kamiokande detector \cite{KSHirata}.

Before discussing the numerical results in detail,  we will try to
understand qualitatively how the $E_{\nu}$ dependence shown in
Fig. 1 naturally arises in RSFP. The argument goes as follows.
For smaller $E_{\nu}$ the level crossing between
$\nu_{e_L}$ and $\overline{\nu}_{\mu_R}$ is
not possible,  and $P_{\nu} \sim 1$. For intermediate $E_{\nu}$ the
level crossing is realized and (as far as $B$ is of suitable strengths)
enough depletion of $\nu_{e_L}$ becomes possible,  i.e. $P_{\nu} \sim 0$.
So far the behavior of $P_{\nu}$ is similar to that in the MSW solution
with relatively small $\theta$. An essential deviation from the MSW case
becomes manifest when we consider the limit of large $E_{\nu}$. In the MSW,
in this limit the transition $\nu_{e_L} \rightarrow {\nu}_{\mu_L}$
tends to be non-adiabatic even though $\nu_{e_L}$ experiences
the level crossing,  and $P_{\nu} \rightarrow 1$.
Note that the adiabaticity condition in MSW,

\begin{equation}
\frac{\Delta m^2 \sin^{2} 2 \theta}{E_{\nu} \cos2 \theta} \gg
\frac{d(ln N_e)}{dt}
\end{equation}

\noindent is not satisfied for larger $E_{\nu}$. The
situation is completely different in RSFP,  since the $E_{\nu}$ dependence
of the argument of the Landau-Zener probability \cite{Parke},  a factor
to judge the adiabaticity of the transition,  is just opposite to the MSW
case. Namely,  the adiabaticity condition in RSFP,

\begin{equation}
\frac{(\mu B)^2
E_{\nu}}{\Delta m^2} \gg \frac{d(ln N_e)}{dt},
\end{equation}

\noindent is well satisfied for large $E_{\nu}$.
In Eq. $(1)$ we realize that when $E_{\nu}$ gets large the off-diagonal
element $\mu B$ in the $2 \times 2$ ``Hamiltonian" matrix becomes
relatively important,  and a ``maximal mixing" between $\nu_{e_L}$ and
$\overline{\nu}_{\mu_R}$ (corresponding to $\theta
\simeq \frac{\pi}{4}$ in the flavor mixing case) takes place.
This is why
$P_{\nu}$ approaches to $\frac{1}{2}$ in the large $E_{\nu}$
limit,  as is seen in Fig. 1.

Let us also point out a sort of scaling property. Fig. 2 shows
profiles of $P_{\nu}$ as a function of $E_{\nu}/\Delta m^2$ for a few
typical values of $B$ with $\mu = 10^{-11} \mu_B$. From this figure
we learn that when $B$ scales as $B \rightarrow \lambda B$,  approximately the
same $P_{\nu}$ is obtained once another parameter also scales as
$(E_{\nu}/\Delta m^2) \rightarrow (1/\lambda^2)(E_{\nu}/\Delta m^2)$,
for a suitable range of $B$. This means that almost the same
$P_{\nu}$ profile is obtained under the change of parameters,
$B \rightarrow \lambda B$ and $\Delta m^2 \rightarrow
\lambda^2 \Delta m^2$. It is interesting to note
that this scaling leaves the argument of the Landau-Zener probability,
$(\mu B)^2 E_{\nu}/(\Delta m^2(d ln N_e/dt))$,  unchanged. Thus $P_{\nu}$
is roughly controlled by the adiabaticity of the transition.

What the scaling property suggests is that the allowed region in the
$(B,\  \Delta m^2)$ plane is roughly along a line
$\Delta m^2/ B^2 \simeq $ constant,  as has been confirmed by the
explicit numerical calculations of the allowed range,  shown in Fig. 3
(assuming $\mu = 10^{-11} \mu_B$). Fig. 3 presents our numerical results
of the allowed regions,
in order to account for four experiments
simultaneously at 99\%,  95\%,  and 90\% confidence levels. Though
the allowed regions in Fig. 3 are rather broad,  too small or too large
magnetic field $B$ is not favored. The reason is that too small $B$ makes
the transition non-adiabatic,  and too large $B$ leads to a uniform
depletion of solar neutrinos by a factor $1/2$ (though the contribution
from the $\overline{\nu}_{\mu_R} e $ scattering causes a little
bit higher event rate in Kamiokande). Thus the allowed region of the
parameters are (at $99 \%$ C.L.),

\noindent $ 25 \leq B \leq 130 \  \ ({\rm kG})$ ,

\noindent $ 7 \times 10^{-9} \leq \Delta m^2 \leq 2 \times 10^{-7}
\ \ ({\rm eV}^2) $.

In the above discussions we assumed a uniform magnetic field $B$,
though it is most probably unrealistic. The reason was twofold.
First,  our knowledge of the magnetic field profile,  especially
in the inner zone,  is very poor. Secondly,  we aimed to extract
as much as possible the genuine feature of RSFP,  irrespectively of
specific choices of $B$ profile,  although the specific choices of
$B$ have been discussed to be helpful in reconciling the three
solar neutrino data in the literature \cite{Akhmedov2,Plido}.
When we consider the possible time-dependence of the event rates,  however,
the assumption of time dependent but uniform $B$ may be too
simple-minded and may lead to a ``really" unrealistic consequence.
This is because the $pp$ neutrinos detected in Ga experiment with
relatively lower energies experience the level crossing at
the inner zone for the values of $\Delta m^2$ inside
the allowed region of Fig. 3,  while the crossing points of
$^8$B and $^7$Be
neutrinos are mostly in the convective zone.
We should note that it is,  at least,  a general consensus that the
magnetic field $B$ varies periodically only in the convective zone,
and the
strengths of $B$ in the two zones can be quite different,  in general.

We therefore will finally analyze in some detail the consequences
of taking
different magnitudes of magnetic fields in convective and
inner zones,  denoted by $B_c$ and $B_i$,  respectively. Only
two extreme cases are considered here as the choices of $B_i$,
i.e.,  (a) $B_i=0$,  and (b) $B_i=10^{3} \ ({\rm kG})$.
In these two  typical cases
the $E_{\nu}/\Delta m^2$ dependence of $P_{\nu}$ have been
shown in Fig. 4,  for several values of $B_c$. Fig. 4 clearly shows
that the lower energy part is greatly affected by the choice of
$B_i$,  since the level crossings of the lower energy neutrinos
happen in the inner zone; when $B_i=0$ (case (a)),  the lower energy
part has almost no depletion,  while when $B_i=10^3 \ ({\rm kG})$ (case (b)),
the lower energy part gets strong suppressions. We have again
calculated the allowed regions in $(B_c,\  \Delta m^2)$ plane to
reconcile all solar neutrino
data at $95 \% $ C.L.  for cases (a) and (b),  whose results have
been demonstrated in Fig. 5. (The region (c) is for the case $B_i=B_c$, and
has already appeared in Fig. 3,  but has been included for the
sake of comparison.)
As is seen in this figure, in the case (a) there scarcely remains the allowed
region only when $B$ is rather strong. This is because the neutrino
oscillation is possible only in the convective zone ($B_i=0$), and
when $B$ is strong we cannot expect the $E_\nu$ dependence of the survival
probability, just as in the OVV scenario.
In the figure in case (b) very small $\Delta m^2$ is favored.
This is because unless $\Delta m^2$ is very small lower energy
neutrinos,  detectable at the Ga experiment,  suffer from too strong
depletion,  as is clear in Fig. 4 for case (b),  which contradicts
with the data. Though all of three cases have their own allowed regions,
each case has different $\chi^2$ minimum:
$\chi^2_{min}= 2.0, 3.8$ and 3.0 for 2 degrees of freedom
(4 data $-$ 2 parameters ) for
 cases (a),  (b),  and (c),  respectively. Thus the case (b)
may not be preferred as a possible solution, and the
case (c) seems to be the most reasonable solution.

How about the time variations of the event rates?
Instead of considering the time variations themselves we will discuss the
variations of the solar neutrino event rates as the functions of the
convective zone magnetic field $B_c$.
Though the average event rates can be most reasonably explained
in case (c),
the assumed relation $B_i=B_c$ should be understood to
hold only in average,  and $B_c$ should be regarded to modulate around the
average value. We therefore choose $B_i$ to be the central values of $B$ in
the allowed region of Fig. 3 (or in Fig. 5 for case (c)) for a few
representative
values of $\Delta m^2$ in the allowed region,  i.e.
$B_i=30,  60, $ and $90 \ ({\rm kG})$ for
$\Delta m^2=10^{-8},  4\times 10^{-8}, $ and $10^{-7} \ ({\rm eV}^2)$,
respectively,  and the variations of the event rates have been plotted as the
functions of $B_c$ in Fig. 6
for these choices of $B_i$ and $\Delta m^2$.
In Fig. 6, it is hard to find out a reasonable parameter choice, which
clearly shows the claimed time dependence of the data.
We have also
calculated the
variations for other cases (b) and (c) as well,  but no better
result has been obtained.

As a summary,  we have argued and have shown by explicit numerical
calculations that all three types of solar neutrino experimental data
on their capture rates can be naturally reconciled by
the Resonant Spin-Flavor Precession (RSFP) scenario. This is
mainly because the $E_{\nu}$ (neutrino energy) dependence of the
$\nu_e$ survival probability $P_{\nu}$,  specific in RSFP,  is
suitable for reconciling the data. Because of this property,  there
appears a rather broad allowed region in the $(B,\  \Delta m^2)$
plane ($B$: strength of magnetic field),  which provides a reasonably
good fit to the three experimental data. The good fit is available
even for a simple-minded assumption of a uniform $B$. We have,  in
addition,  analyzed the consequences of taking different $B$'s in
convective and inner zones: $B_c,  B_i$ respectively.
We have found that if $B_i$ is very weak a there scarcely remains the
allowed region,
while too strong $B_i$ seems to be not preferred. As for the time
variations of the event rates of three types of experiments,  to get
reasonable behaviors seems to be non-trivial,  and
in our rather restricted  choice of parameter set we have not succeeded
in it.
Of course,  various additional
possibilities will be opened if we adopt more elaborate
$B$ profiles \cite{Akhmedov2,Plido},  or if we include
a generation mixing in addition to RSFP
(see Ref. \cite{HNunokawa}).

\vglue 0.5cm
\noindent {\large{\bf Figure Captions}}\par
\begin{tabbing}
\= artiffi    \=                     \kill
\>Fig. 1:  \>\ \  The conceptualized $E_{\nu}/
\Delta m^2$ dependence of \\
\>        \> the $\nu_e$ survival probability
$P(\nu_e \rightarrow \nu_e)$ in RSFP\\
\>        \> \\
\>Fig. 2:  \> \ \ The profiles of $P_{\nu}$ as the
functions of   $E_{\nu}/\Delta m^2$  \\
\>        \> for $B$ = 10,  30, 50 and 100 \ kG respectively \\
\>        \> with $\mu = 10^{-11}\mu_B$ assumed.     \\
\>        \> \\
\>Fig. 3:  \> \ \ The allowed regions to account for the
data of the three \\
\>        \> types solar neutrino experiments simultaneously,  \\
\>        \> at $99,  95, $ and $90 \%$ C.L.   \\
\>        \> \\
\>Fig. 4:  \> \ \ The $E_{\nu}/\Delta m^2$ dependence
of $P_{\nu}$ for \\
\>        \> two different choices of the inner
zone magnetic field \\
\>        \> $B_i$,  i.e.  (a)$B_i=0$,  and (b) $B_i=10^3
\ ({\rm kG})$,  each \\
\>        \> for $B_c$ = 10, 30, 50,  and  100 \  kG.  \\
\>        \> \\
\>Fig. 5:  \> \ \ The allowed regions in
$(B_c,  \Delta m^2)$ plane (95\% C.L.), similar\\
\>        \>  to Fig. 3,  but now assuming
two typical cases for the strength\\
\>        \>  of $B_i$, i.e.  $B_i = 0$
(region (a), disconnected 5 black areas)\\
\>        \>  and $B_i = 10^3$ \ kG (region (b)).
The region (c) is the same as\\
\>        \> in Fig. 3 but has been included for comparison. \\
\>        \>  \\
\>Fig. 6:  \> \ \ The variations of event rates
of three types of experiments \\
\>        \> as the functions of $B_c$
for three choices of remaining \\
\>        \> parameters,  i.e. (i)\ $B_i=30 \ ({\rm kG}),\
 \Delta m^2=10^{-8} \ ({\rm eV}^2)$,  \\
\>        \> (ii)\ $B_i=60 \ ({\rm kG}),\
\Delta m^2=4\times 10^{-8} \ ({\rm eV}^2)$,  and \\
\>        \> (iii)\ $B_i=90 \ ({\rm kG}),\
\Delta m^2= 10^{-7} \ ({\rm eV}^2)$. \\
\>        \>  \\
\end{tabbing}
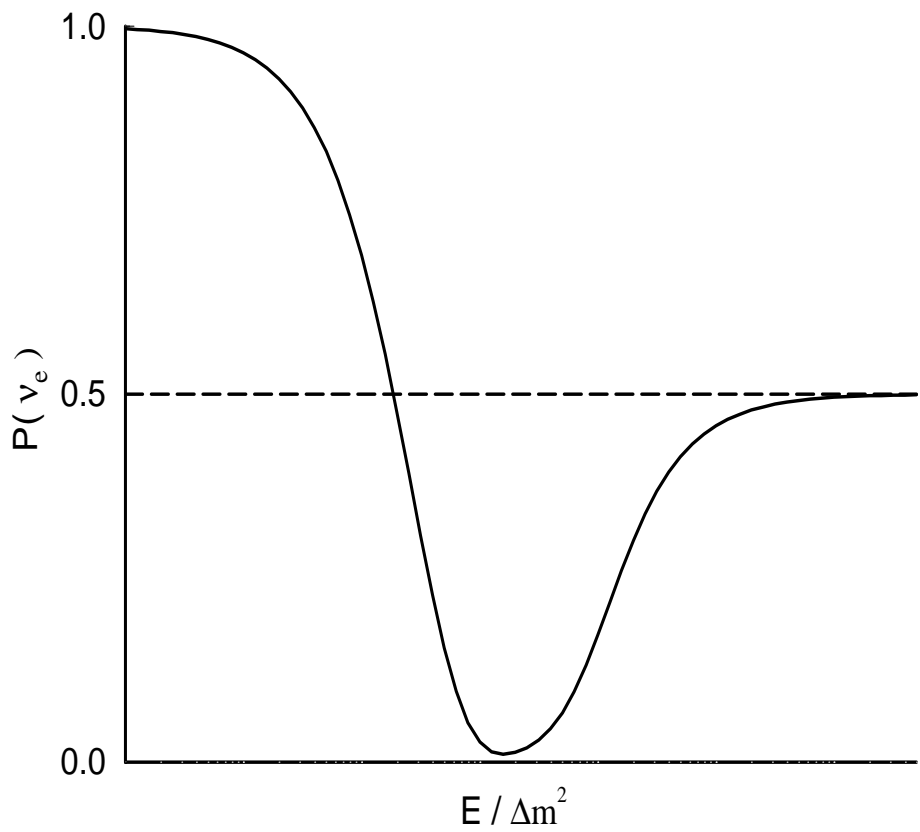
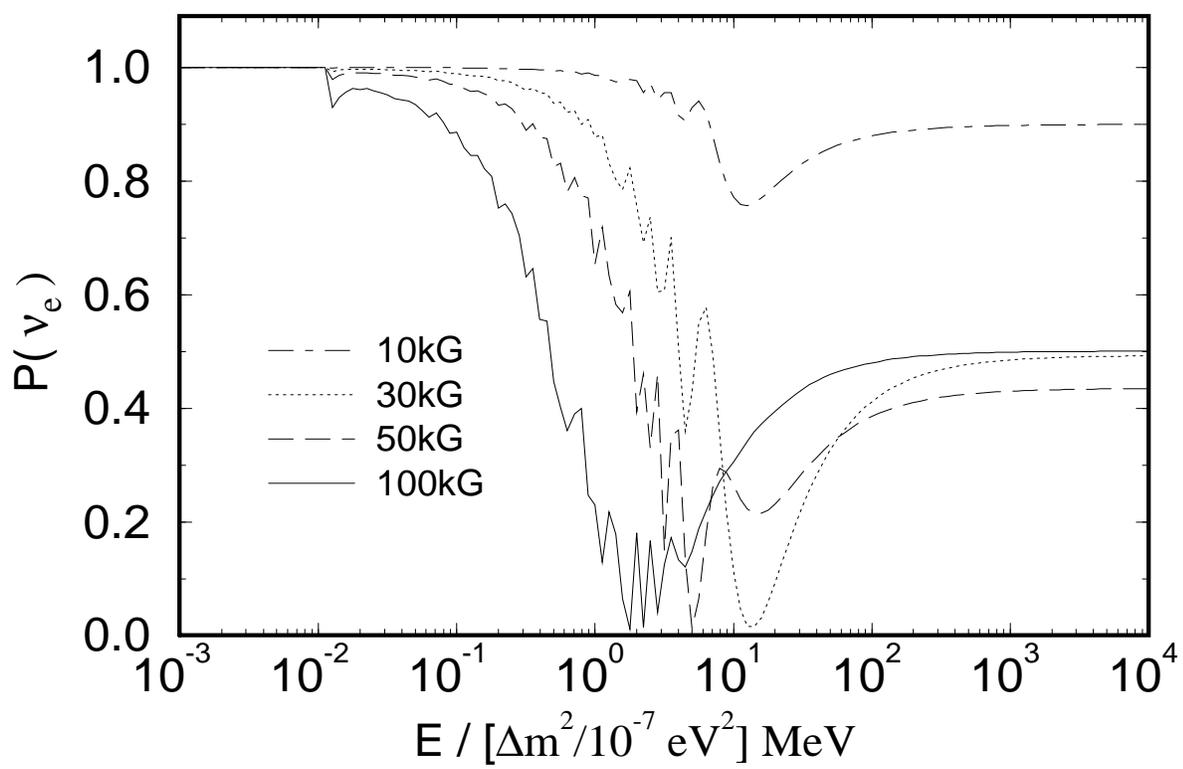
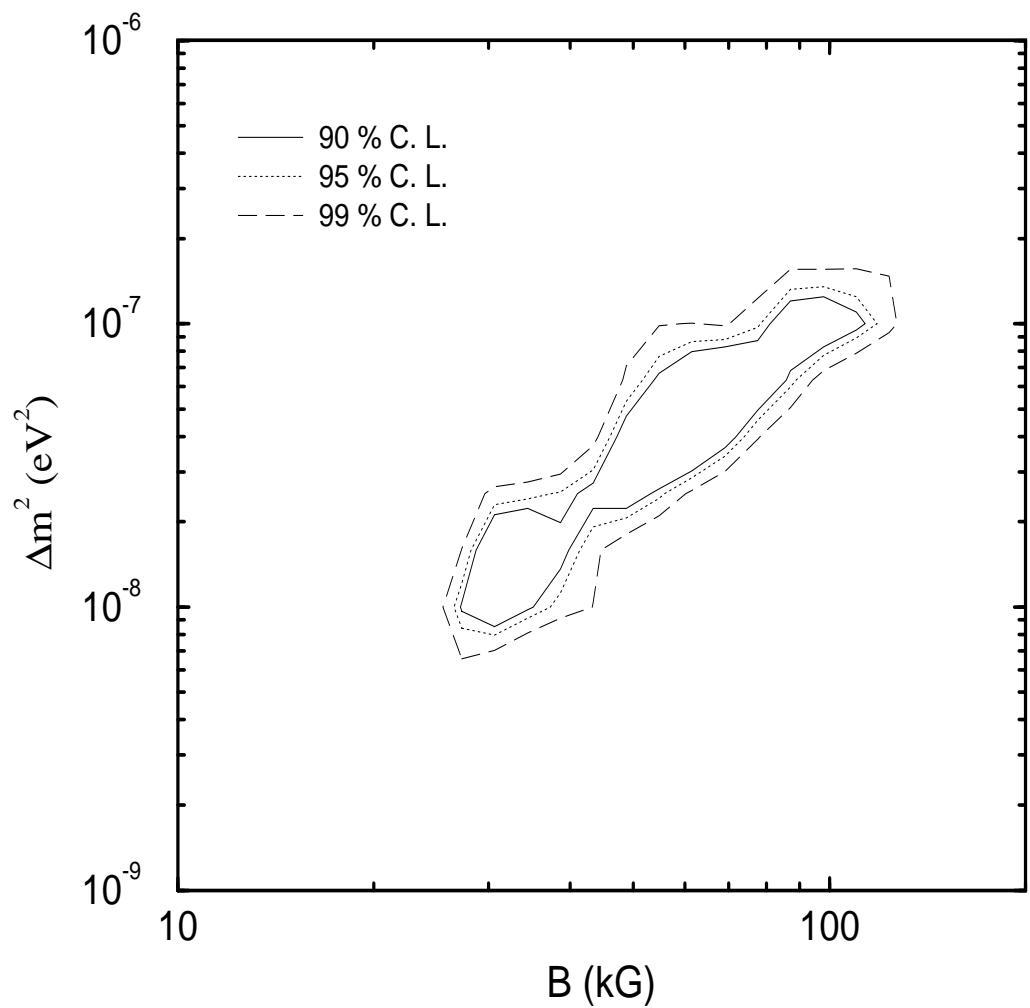
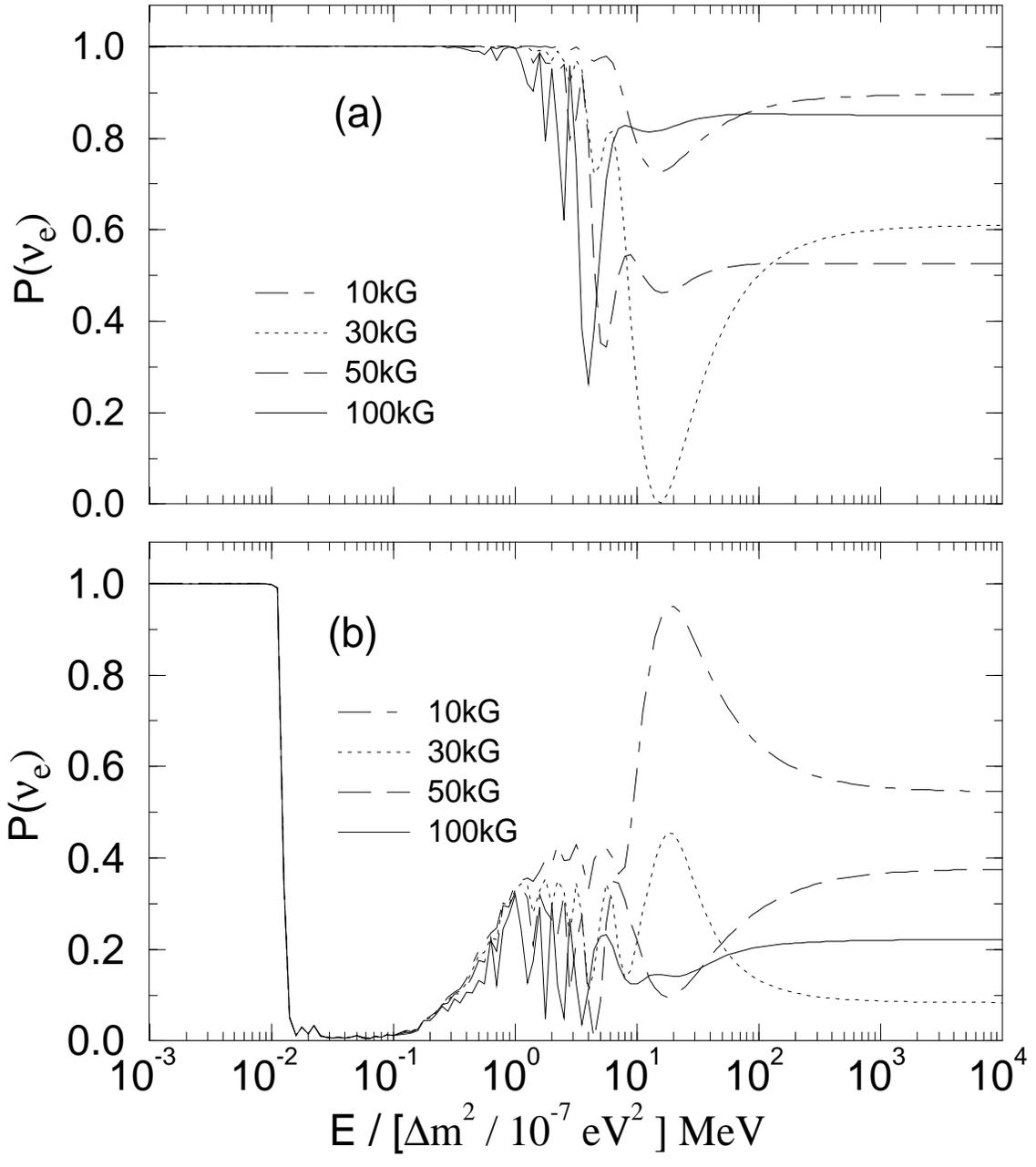
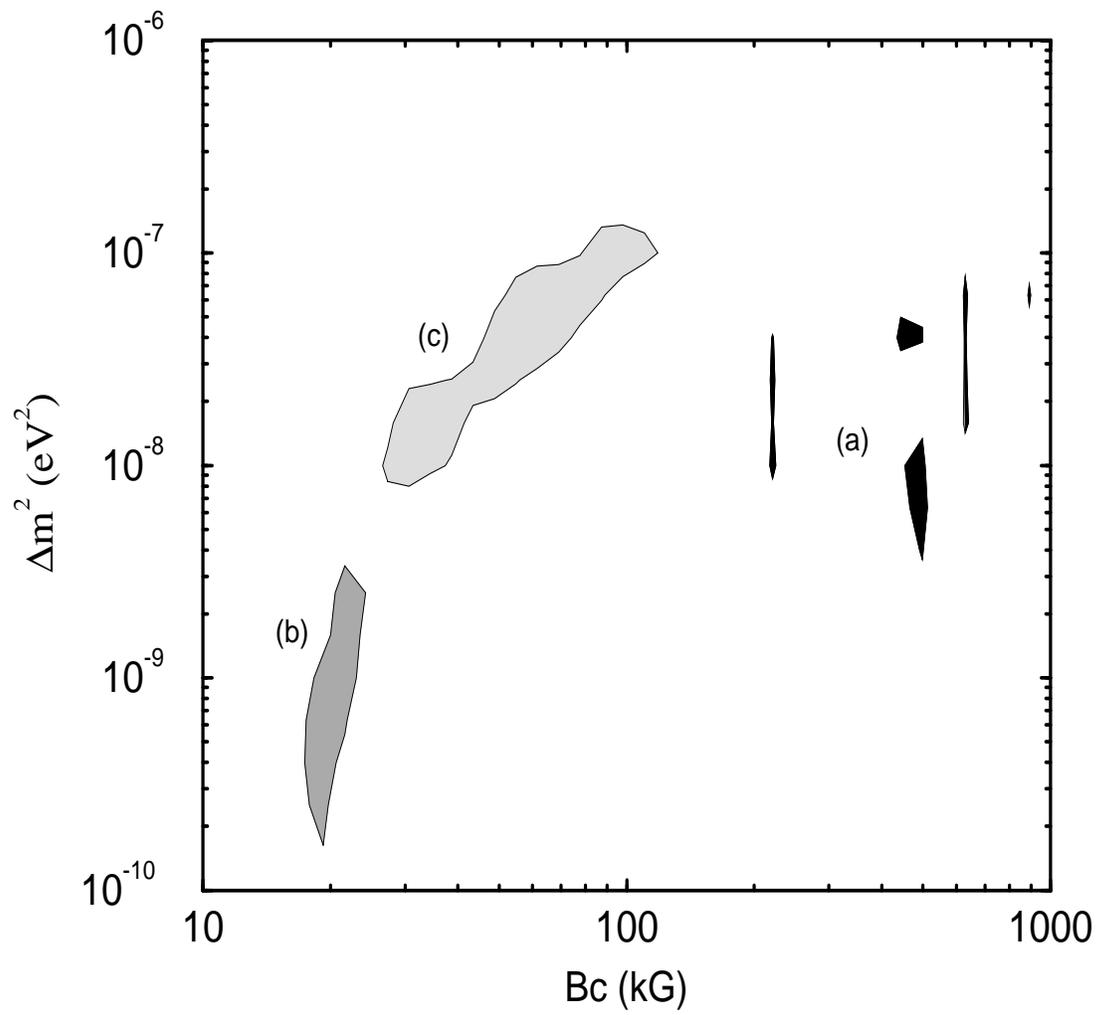
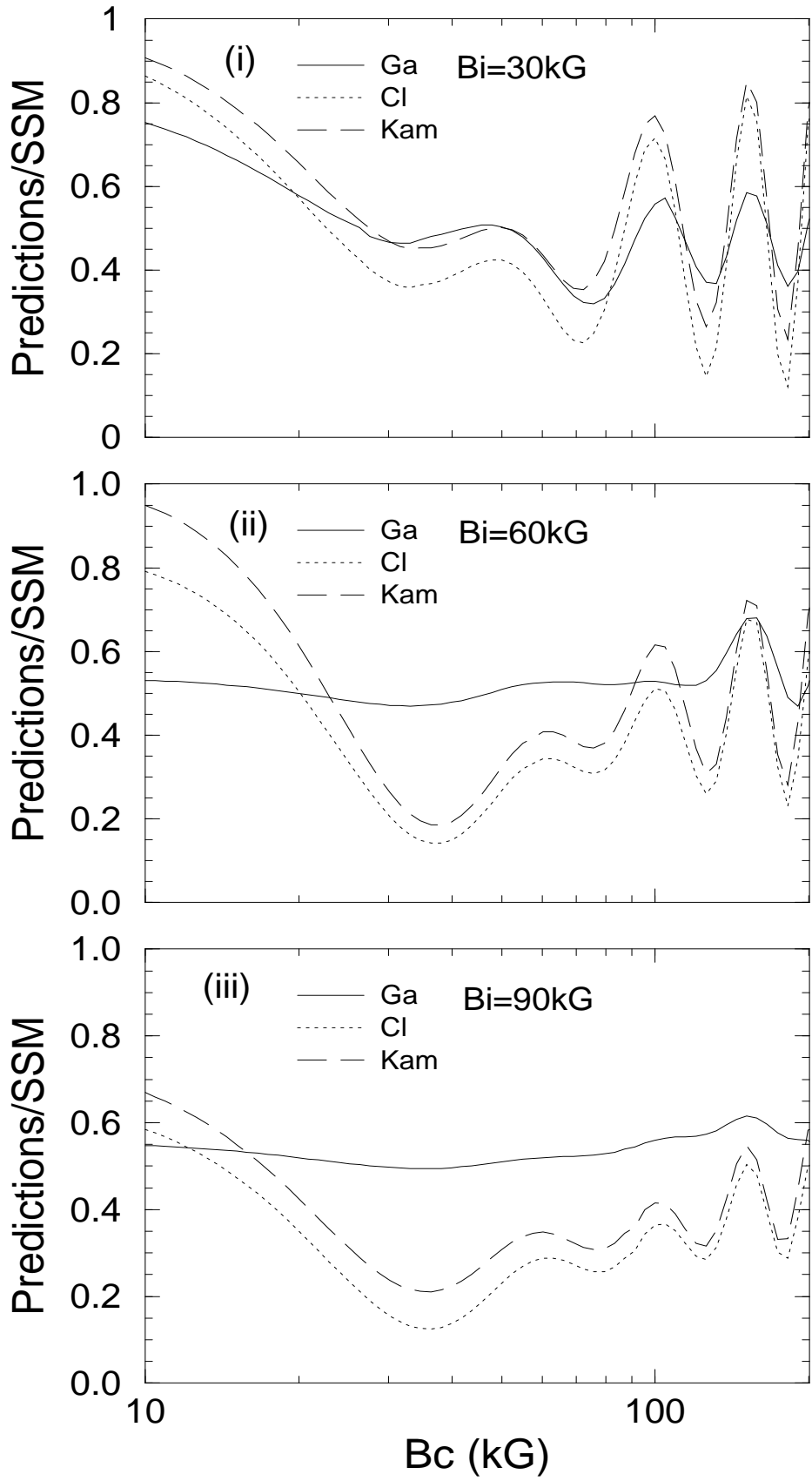
\end{document}